\documentclass[aps,prd,twocolumn,preprintnumbers,superscriptaddress,dblfloatfix,nofootinbib]{revtex4-1}
\usepackage[utf8]{inputenc}
\usepackage{lmodern}
\usepackage{amsmath,amssymb,mhchem}
\usepackage{graphicx}
\usepackage{url}
\usepackage{color}
\usepackage{enumitem}
\usepackage{subfigure}
\usepackage[dvipsnames]{xcolor}
\usepackage[colorlinks=true,breaklinks=true]{hyperref}
\hypersetup{allcolors=[rgb]{0.0 0.0 0.6},linkcolor=[rgb]{0.75 0.05 0.05}}
\usepackage{bm}
\usepackage{epsfig}
\usepackage{amsmath}
\usepackage{amssymb}
\usepackage{slashed}
\usepackage{color}
\usepackage{accents}
\usepackage{url}
\usepackage[dvipsnames]{xcolor}
\usepackage{cancel}
\usepackage[normalem]{ulem} 
\usepackage[colorinlistoftodos]{todonotes}
\usepackage{multirow}

\renewcommand\[{\left[}

\newcommand{\exclude}[1]{}
\newcommand{\Sec}[1]{Sec.~\ref{#1}}

\newcommand{\wh}{\widehat}
\newcommand{\neut}{{\tilde{\chi}^0_1}}

\begin{document}


\title{Searching for light long-lived neutralinos at Super-Kamiokande}

\author{Pablo Candia}
\email{pablo.candiadasilva@manchester.ac.uk}
\affiliation{Consortium for Fundamental Physics, School of Physics and Astronomy, University of Manchester, Manchester, M13
  9PL, United Kingdom}

\author{Giovanna Cottin}
\email{giovanna.cottin@uai.cl}
\affiliation{Departamento  de  Ciencias,  Facultad  de  Artes  Liberales, \\ Universidad  Adolfo  Ib\'a\~nez,  Diagonal  Las  Torres  2640,  Santiago,  Chile}

\author{Andr\'es M\'endez}
\email{aimendez@uc.cl}
\affiliation{Instituto  de  F\'isica,  Pontificia  Universidad  Cat\'olica  de  Chile,  Avenida  Vicu\~na  Mackenna  4860, Santiago, Chile}

\author{V\'ictor Mu\~noz} 
\email{victor.manuel.munoz@ific.uv.es}
\affiliation{Instituto de F\'isica Corpuscular (IFIC), CSIC-Universitat de Valencia, \\ Apartado de Correos 22085, E-46071, Spain}

\date{\today}

\begin{abstract}
Light neutralinos could be copiously produced from the decays of mesons generated in cosmic-ray air showers. These neutralinos can be long-lived particles in the context of R-parity violating (RPV) supersymmetric models, implying that they could be capable of reaching the surface of the earth and decay within the instrumental volume of large neutrino detectors. In this article, we use atmospheric neutrino data from the Super-Kamiokande experiment to derive novel constraints for the RPV couplings involved in the production of long-lived light neutralinos from the decays of charged $D$-mesons and kaons. Our results highlight the potential of neutrino detectors to search for long-lived particles, by demonstrating that it is possible to explore regions of parameter space that are not yet constrained by any fixed-target nor collider experiments.  \end{abstract}

\maketitle
{
  \hypersetup{linkcolor=black}
  \tableofcontents
}


\section{Introduction}

The discovery of the Higgs boson at the Large Hadron Collider (LHC) in 2012~\cite{Aad:2012tfa,Chatrchyan:2012ufa} provides not only conclusive evidence of the Standard Model (SM), but also consolidates the hierarchy problem as one of the main theoretical puzzles in modern physics~\cite{Giudice:2013yca}. In this context, supersymmetry (SUSY) remains as one of the most compelling possibilities to address this problem~\cite{Martin:1997ns,2017Prama..89...53B}. At the same time, supersymmetry also provides a rich and complex phenomenology which has lead to an intensive search program at collider experiments~\cite{Canepa:2019hph}.\\
Conventional SUSY theories assume a discrete symmetry called R-parity, which avoids conflict with experimental data on the non-observation of baryon and lepton number violating processes, such as proton decay \cite{Font:1989ai} and neutrinoless double beta decay \cite{PhysRevD.34.3457}.
Within the context of R-parity conserving SUSY theories, the lightest neutralino as the lightest supersymmetric particle (LSP) provides a natural candidate for fermionic dark matter because of its stability and lack of electromagnetic interactions (see \cite{PhysRevLett.50.1419,Ellis:1983ew} for seminal articles and \cite{Roszkowski:2017nbc} for a review). Yet, it is possible to assume R-parity violation (RPV)~\cite{Barbier:2004ez} while respecting the bounds on the proton lifetime, as long as baryon number or lepton number are preserved. In such a case, the neutralino LSP is no longer stable and can decay into SM particles. The smallness of the R-parity violating couplings can make the decay macroscopic, making the neutralino, $\tilde{\chi}^{0}_{1}$, a long-lived particle (LLP)~\cite{Graham:2012th}. This implies that neutralinos can leave a variety of exotic signatures at colliders such as a displaced vertex (reviews on possible long-lived particle signals can be found in \cite{Lee:2018pag,Alimena:2019zri}). On the other hand, unlike the strong interacting sparticles whose masses have a lower bound around 1 TeV \cite{Aaboud:2018doq,Sirunyan:2017nyt,Sirunyan:2019mbp,Sirunyan:2019ctn,Aad:2020nyj,10.1093/ptep/ptaa104,Khachatryan:2016iqn,Aaboud:2017opj,Khachatryan:2016ycy}, the neutralino mass is less constrained, and can in principle be massless \cite{Gogoladze:2002xp}. For a fraction of the R-parity violating model parameter space, the ATLAS and CMS experimental collaborations at the LHC have searched for a long-lived $\tilde{\chi}^{0}_{1}$ with a mass $\mathcal{O}(100)$ GeV with a displaced vertex signature~\cite{Aad:2015rba,Aad:2019tcc,Khachatryan:2016unx}, with null results. The most up-to-date constraint comes from ATLAS with leptonic displaced decays, excluding long-lived neutralinos between $50-500$ GeV~\cite{Aad:2019tcc}. 

Phenomenological prospects demonstrate that in extended regions in the R-parity violating couplings -- leading to decays with different leptonic or hadronic final state particles -- and lighter masses below 100 GeV, a long-lived $\tilde{\chi}^{0}_{1}$ has the potential to be discovered at dedicated LLP experimental facilities that could operate at the LHC, such as FASER, MATHUSLA, CODEX-b and AL3X~\cite{Helo:2018qej,Dercks:2018eua,Dercks:2018wum,deVries:2015mfw,Dey:2020juy,Wang:2019orr}, future colliders at the intensity frontier~\cite{Wang:2019orr} or beam-dump experiments as SHiP~\cite{deVries:2015mfw}. In particular, the absence of experimental constraints in the range from few MeV to $\mathcal{O}$(1) GeV makes it a good candidate to be studied in scenarios were they can be produced from the mesons that are abundantly created in both colliders and cosmic-ray air showers.\\

 So far, prospects for light, long-lived neutralinos have been performed from the decays of $D$ and $B$ mesons in references \cite{Alekhin:2015byh,Gorbunov:2015mba,deVries:2015mfw,Dercks:2018eua,Dreiner:2020qbi}, from $Z$ boson decays in \cite{Helo:2018qej,Dercks:2018wum,Wang:2019xvx,Wang:2019orr}, and from the decay of $\tau$ leptons in Belle-II \cite{Dey:2020juy}. Most of these searches are based in experiments that are either in construction like FASER \cite{Ariga:2018uku}, or even in earlier stages as they are subject to funding/approval (\textit{e.g.} SHiP \cite{Buonaura:2017gos}, MATHUSLA \cite{Alpigiani:2020tva} and others \cite{Aielli:2019ivi,Bauer:2019vqk}). In contrast to this situation, large Cherenkov based neutrino detectors such as IceCube \cite{IceCube:2008qbc} or Super-Kamiokande (SK) \cite{Super-Kamiokande:2002weg} are already built and have been taking data for years, which can be used to search for LLPs, as the decay of these particles would generate a signal that is indistinguishable from the Cherenkov radiation measured in association with the neutrino interactions in the medium.\\

Searches for long-lived particles at large neutrino detectors were considered in the literature in references~\cite{Kusenko:2004qc,Asaka:2012hc,Masip:2014xna,Arguelles:2019ziu,Coloma:2019htx,Meighen-Berger:2020eun}. Of these studies, reference \cite{Arguelles:2019ziu} used public data from IceCube and Super-Kamiokande to search for long-lived particles produced in atmospheric showers, using detailed numerical simulations. The results found demonstrated that the atmospheric neutrino data from the Super-Kamiokande experiment can be used to place stringent constraints to models predicting LLPs (see also reference \cite{Coloma:2019htx}). The main reason for this is because the atmospheric neutrino data focused on the low energy regime, where the flux of the atmospheric particles peaks. In this work we build upon the general results of \cite{Arguelles:2019ziu}, and apply the same strategy to search for neutralinos produced from meson decays, including an appropriate treatment of the uncertainties originated from the hadronic interaction models used to simulate the production of mesons in the shower.

We consider the possibility of searching for neutralinos produced in the decays of $D$-mesons and kaons that are generated in the sky, when highly energetic cosmic rays collide with the upper layers of the atmosphere. The former process allows us to compare with the previously mentioned studies in the literature, particularly references \cite{Dercks:2018eua} and \cite{deVries:2015mfw}, where the sensitivity reach at collider experiments was estimated for selected benchmarks. On the other hand, neutralino production from kaon decay was considered long before in the context of spontaneously broken supersymmetry~\cite{Gaillard:1982rw}. In this work, we explore a novel RPV channel through kaon production that is particularly well suited for our setup because the production of kaons in the atmosphere greatly exceeds the one from $D$-mesons~\cite{Fedynitch:2018cbl}.

The rest of the paper is organized as follows. In \Sec{sec:CR} we summarize the relevant phenomenological aspects of the RPV theory, and describe the neutralino production from the decay of pseudoscalar mesons generated in cosmic-ray air showers. In \Sec{sec:SK}, we describe the expected signal from the visible decay of the neutralino in the Super-Kamiokande detector. The main results obtained for the two benchmark scenarios considered in this work are presented in section \Sec{sec:results}, along with the current best constraints from previous studies. Finally, we draw our main conclusions in section \Sec{sec:conclusion}.

\section{Long-lived light neutralinos in cosmic-ray air showers}\label{sec:CR}

\subsection{Neutralinos in RPV}
\label{sec:RPV}
The simplest realistic realization of a supersymmetric model is the so-called minimal supersymmetric standard model (MSSM) \cite{Haber:1984rc}. As mentioned above, in this model a $Z_2$ symmetry called R-parity is imposed to avoid proton decay. However, it is still possible to break R-parity and have a stable proton by imposing a different discrete symmetry like the baryon triality $B_3$ symmetry \cite{Dreiner:2005rd}. Without imposing R-parity, the most general Lagrangian that respects gauge and space-time symmetries contains the term \cite{PhysRevD.26.287}
\begin{equation}
\mathcal{L}\supset \lambda'_{ijk} \wh{L}_i \wh{Q}_j \wh{D}^c_k,
\end{equation}
where the hatted symbols denote gauge multiplets of superfields that include leptons $\ell_i$ and sleptons $\tilde{\ell}_i$ in $\wh{L}_i$, left-handed quarks $q_{jL}$ and squarks $\tilde{q}_{jL}$ in $\wh{Q}_j$, and right-handed antiquarks $\overline{d}_{kR}$ and antisquarks $\tilde{d}^*_{kR}$ in $\wh{D}^c_k$. The family indices $i$, $j$ and $k$ run from 1 to 3, leading to 27 independent parameters. This semileptonic operator violates lepton number by one unit, and can generate contributions to the mass and magnetic moment of neutrinos \cite{Bhattacharyya:1999tv}, neutrinoless double beta decay \cite{Babu:1995vh,Hirsch:1995ek}, and in the parameter space of our interest, neutralino production from meson decays \cite{Choudhury:1995pj,Choudhury:1999tn,Dedes:2001zia}. Comprehensive reviews of RPV and its phenomenological implications can be found in \cite{Dreiner:1997uz,Bednyakov:1999bd,Barbier:2004ez,Mohapatra:2015fua}.

\begin{figure*}
\centering
\includegraphics{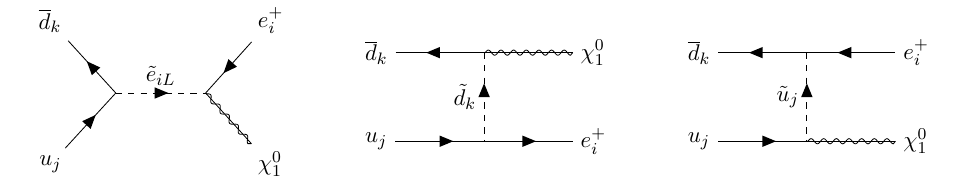}
\caption{Neutralino production diagrams. We display the specific case where production occurs via the decay of a charged meson $M^+_{kj}$ composed of quarks $d^C_k$ and $u_j$. 
For the cases of our interest (see table \ref{tab:benchmark_table}), we set the flavour indices of the initial states in the diagram such that $M^+_{kj}$ corresponds to  $D^+$ $(k=1, j=2)$  for benchmark B1 and $K^+$ $(k=2, j=1)$ for benchmark B2.
}

\label{Fig:Neutralino_production}
\end{figure*}

In this work we focus on the trilinear couplings $\lambda^{\prime}_{121}$ and $\lambda^{\prime}_{112}$, which allow to produce neutralinos from the decays of $D$-mesons and kaons, respectively. These decays are accompanied always with an electron, as detailed in our two benchmarks in table \ref{tab:benchmark_table}. The introduction of RPV parameters also allows neutralino two-body decays into a meson and a lepton.  In our first benchmark, we study neutralinos with masses in the range from the mass of the kaon to the mass of the $D$-meson, while in the second benchmark case we restrict the neutralino mass to the interval between the mass of the pion and the mass of the kaon. Therefore, for the second benchmark, we include the trilinear $\lambda^{\prime}_{111}$, as it is the only coupling associated with pions, and is therefore necessary to make the neutralino decay when its mass is lower than the kaon mass. 
The relevant Feynman diagrams involved in the production of neutralinos from these mesons are shown in figure \ref{Fig:Neutralino_production}.


To calculate the decay rate of the mesons to neutralinos, we closely follow reference \cite{deVries:2015mfw} and consider the effective interactions involving the relevant mesons, leptons and neutralinos. Assuming that the sfermion masses involved are degenerate and large enough so that they can be integrated out, the relevant decay rates are
\begin{align}
\Gamma(D^\pm\rightarrow\neut + e^\pm) = &\frac{\mathcal{K}^{1/2}(m^2_D,m^2_{\neut},m^2_e)}{64\pi m^3_{D}(m_c+m_d)^2}
|G^{S}_{121}|^2 f_{D}^2\nonumber\\
&\times m^4_D(m^2_{D} - m^2_{\neut} - m^2_{e})
\end{align}
and
\begin{align}
\Gamma(K^\pm\rightarrow\neut + e^\pm) = &\frac{\mathcal{K}^{1/2}(m^2_K,m^2_{\neut},m^2_e)}{64\pi m^3_{K}(m_u+m_s)^2}
|G^S_{112}|^2f_{K}^2\nonumber\\
&\times m^4_K(m^2_{K} - m^2_{\neut} - m^2_{e}),
\end{align}
where $\mathcal{K}(x,y,z)\equiv x^{2}+y^{2}+z^{2}-2xy-2xz-2yz$, is the Källén function \cite{Kallen:1964lxa} and $f_D$, $f_K$ are the $D$-meson and kaon decay constants, respectively. The effective couplings $G^S_{ijk}$ are given by \begin{align}
G^{S}_{ijk} \equiv \frac{3g_2}{2\sqrt{2}}\frac{\lambda^{\prime}_{ijk}}{m^2_{\tilde{f}}}\tan\theta_W,\label{eq:scalar_vertex}
\end{align}
where $m_{\tilde{f}}$ represents the common value that we assume for the masses of sleptons and squarks, which are the sfermions involved in neutralino production (details can be found in reference \cite{deVries:2015mfw}). In equation (\ref{eq:scalar_vertex}), $\theta_W\equiv \tan^{-1} g_1/g_2$ is the weak mixing angle \cite{10.1093/ptep/ptaa104}, and $g_1$, $g_2$ are the gauge coupling constants associated to U(1)$_Y$ and SU(2)$_L$, respectively. The symbols $m_K$ and $m_D$ denote the masses of the $K^\pm$ and $D^\pm$ mesons, and $m_c$, $m_d$, $m_u$ and $m_s$ are the masses of the charm, down, up and strange quarks, respectively. In our calculations, we use the values $f_K\simeq 156$ MeV and $f_D\simeq 213$ MeV \cite{10.1093/ptep/ptaa104}.

\subsection{Neutralinos from D-meson and kaon decays in atmospheric showers}\label{sec:mesons}
 
Cosmic rays hitting our atmosphere provide us with a beam of protons (and other species) that is constantly switched on. A single cosmic-ray can produce an extensive cascade of charged particles and radiation called an air shower \cite{Kampert,1960PThPS..16....1F,Galbraith}. Mesons in the shower, including $D$-mesons and kaons, decay to charged leptons and neutrinos, among other particles. The flux of leptons can be measured both at the surface of the earth and in underground experiments, while their careful reconstruction allows to estimate the expected mesonic contributions to the spectrum \cite{Volkova:2001th,PhysRevD.38.85,PhysRevD.39.3532,Lipari:1993hd,Gondolo:1995fq,Fedynitch:2012fs,Fedynitch:2015kcn,Fedynitch:2015zma,Fedynitch:2018cbl}.\\
Similarly as with the case of neutrinos, we simulate the production of light neutralinos in the shower by solving the cascade equation involving only source terms from meson decays \cite{Gondolo:1995fq} 
\begin{align}
\frac{d\Phi_\neut}{dE_{\neut}\,d\Omega\,dX} = \sum_{M}\int d E_{M} \,\frac{1}{\rho \lambda_{M}} \frac{d \Phi_{M}}{ dE_{M} \,d\Omega}\, \frac{d n}{d E_\neut},
\label{eq:production}
\end{align}
where the sum runs over all possible mesons that can decay to neutralinos when a given trilinear coupling $\lambda^{\prime}_{ijk}$ is switched on. In equation (\ref{eq:production}), $\rho$ is the density of the atmosphere at a column depth $X$, and $\lambda_{M} \equiv \gamma_{M}\beta_{M}c\tau_{M} $ is the decay length of the meson, which includes the boost factor $\gamma_{M}\beta_{M}$ and its proper lifetime $\tau_{M}$. The differential production rate of mesons in the shower per unit of solid angle is given by $\frac{d \Phi_{M}}{ dE_{M} \,d\Omega}$. The number of neutralinos with energies between $E_{\neut}$ and $E_{\neut}+dE_{\neut}$ produced in the decay of the meson $M$ is given by  $\frac{d n}{d E_\neut}$. For two-body decays, this last quantity is given by
\begin{align}
\frac{dn}{d E_\neut}= \frac{ \mathrm{Br}(M \rightarrow \neut + e)} 
{ p_M \sqrt{ \mathcal{K} \left(1,\frac{m_{\chi}^2}{m_{M}^{2}}, \frac{m_{e}^2}{m_{M}^2}\right) }
},
\label{eq:distribution}
\end{align}
where $\mathrm{Br}(M \rightarrow \neut + e)$ is the branching fraction of meson decays to neutralinos and $p_{M}$ is the meson momentum.\\
In equation (\ref{eq:production}), both the density of the atmosphere and the differential production rate of mesons are extracted using the Matrix Cascade Equation (\texttt{MCEq}) software package~\cite{Fedynitch:2015zma,Fedynitch:2012fs}. Here, we choose the \texttt{NRLMSISE-00} atmospheric model~\cite{Picone:2002go}, while for the hadronic interaction model we consider the different event generators which are updated with LHC data (see for instance, references \cite{Pierog:2013ria,LHCf:2020hjf}). In particular, we focus on the \texttt{SYBILL-2.3}~\cite{Fedynitch:2018cbl}, \texttt{QGSJET-II-04}~\cite{Ostapchenko:2010vb}, \texttt{EPOS-LHC} ~\cite{Pierog_2009}, and \texttt{DPMJET-III}~\cite{Roesler_2001} models. These models might yield  non-negligible differences in the production rate of mesons and can be a relevant source of uncertainty in our calculations\footnote{Another relevant input in this calculation is the cosmic-ray model chosen for the primary spectra. We checked that the uncertainties from this election are sub-leading in the energy regime considered in this work. We use the Hillas-Gaisser cosmic-ray model \texttt{H3a}~\cite{Gaisser_2012} in our calculations.}.

As an illustration of our numerical simulations, we show in figure \ref{fig:neutralino_production} the production rate for a 0.31 GeV neutralino, at a height $h$ of 15.51 km and with a vertical direction of 25.84º degrees from the zenith, produced in the decay of both $D$-mesons and kaons. The spread of the orange band in the plot covers the most pessimistic and optimist cases given the uncertainties associated with the election of an event generator. Note that there is an important difference in the amount of neutralinos produced from the decay of $D$-mesons and kaons, as the latter are expected to be orders of magnitude more abundant in the atmosphere \cite{Fedynitch:2018cbl}. Finally, it is important to remark that we were not able to assess the uncertainty that pertains the simulations from $D$-mesons, as currently the only updated hadronic interaction model for charmed hadrons is \texttt{SYBILL-2.3} \cite{Fedynitch:2018cbl,Aartsen:2015nss,Albert:2021pwz,Abreu:2020ddv}.\\
The uncertainties are estimated following reference \cite{ArguellesDelgado:2021lek} (see also reference \cite{Kachelriess:2021lpm} for a similar discussion). For a given meson, we calculate the total expected flux $\frac{d\phi}{dE}$ at the surface of the earth with different hadronic models. In order to do so, we let the mesons propagate through the atmosphere without decaying, which can be accomplished by switching off their decay in \texttt{MCEq}. The impact of the variation in the meson production rate for different models is then quantified with respect to a benchmark model, which we take here
to be the \texttt{SYBILL-2.3} model, by defining the ratio $\Delta_{\texttt{j}}(M)$ between the meson flux predicted by the event generators that are being compared:
\begin{figure}
\begin{center}
\includegraphics[width=0.99\columnwidth]{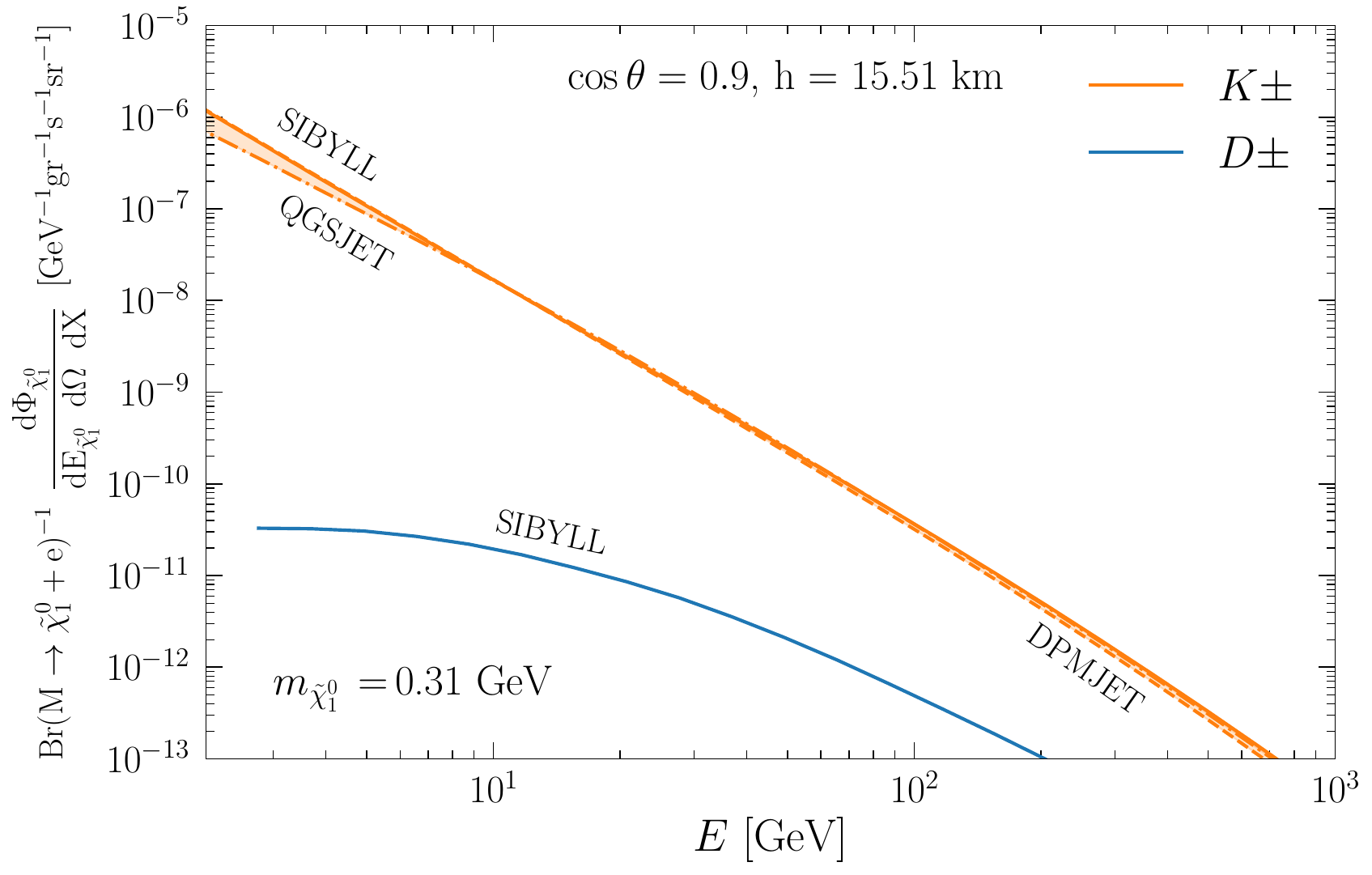} 
\caption{
\textbf{\textit{Neutralino Production Rate.}}
Energy spectrum of the atmospheric production rate of light neutralinos from charged $D$-mesons (blue curve), and kaons (orange band). The spread of the orange band reflects the difference in the production originated from the election of different hadronic event generators; \texttt{SIBYLL} (solid line), \texttt{QGSJET} (dot-dashed line), and \texttt{DPMJET} (dashed line). We choose a representative mass for the neutralino with a value of 0.31 GeV. The angular direction is fixed at $\cos\theta=0.9$, while the height was chosen around 15 km, where a maximum production rate is expected. }
\label{fig:neutralino_production}
\end{center}
\end{figure}
\begin{equation}
\Delta_{\texttt{j}}(M) = \frac{\int_{E_{\rm min}}^{\Lambda} dE\, \frac{d\phi_{\texttt{BM}}}{dE}}{\int_{E_{\rm min}}^{\Lambda} dE\, \frac{d\phi_{\texttt{j}}}{dE}} ,
\label{eq:uncertain}
\end{equation}
where $M$ is the meson of interest, \texttt{j} is the index used to denote the model that is being compared with the benchmark model \texttt{BM}, $E_{\rm min}= 1.6 $ GeV is the minimum energy available in \texttt{MCEq}, and $\Lambda = 10^{3}$ GeV is the upper energy cutoff that we use in order to obtain the neutralino production rate from a given parent meson. Table~\ref{tab:uncert_val} shows the $\Delta$ coefficients obtained for the production of kaons with different event generators.
The uncertainty as quantified by equation (\ref{eq:uncertain}) reaches a maximum of around 66$\%$ for \texttt{QGSJET}. We calculate the neutralino production rate using different hadronic interaction models, and assess their impact on the region of parameter space that can be probed. We stress that dedicated efforts are needed in order to reduce the uncertainties involved in meson production in the forward direction, a problem that is also crucial for precise modeling of neutrino fluxes at the LHC \cite{Ismail:2020yqc,Kling:2021gos,Bai:2020ukz}.
\begin{table}[t]
\centering
\begin{tabular}{||c||c||} 
  \hline
 \textbf{Model}  &    $\Delta_{\texttt{j}}(K^{\pm})$\\ 
\hline
 \hline
 \texttt{DPMJET}&  1.163 \\
 \texttt{EPOS-LHC}  & 1.116 \\
 \texttt{QGSJET} & 1.660 \\
 \hline
\end{tabular}
\caption{\textbf{\textit{Comparison of the meson flux using different hadronic interaction models.}} Numerical values of $\Delta$, the relative integrated meson flux with respect to \texttt{SYBILL}, for the different hadronic interaction models considered in this work.}
\label{tab:uncert_val}
\end{table}
\section{Neutralino signals at Super-Kamiokande} \label{sec:SK}

After light neutralinos are produced from the decay of mesons in the atmospheric shower, they decay to SM particles as they propagate. We assume that all the particles in the decay chain are highly boosted, and hence approximately collinear in their trajectories. The degree of attenuation of the flux that arrives at the detector depends on the relation between the lifetime of the neutralino and the distance it travels before reaching the detector. In particular, it is expected that the detector signal from up-going events ($i.e.$ the neutralinos that arrive from below the detector), will be suppressed in comparison with down-going events ($i.e.$ the neutralinos that reach the detector from above). Assuming that the production rate of neutralinos is symmetric with respect to the azimuthal component, the expected differential flux is obtained by integrating over the column depth $X$, according to
\begin{equation}
\frac{d\Phi_\neut}{dE_\neut\,d\cos\theta} = 2\pi \int dX \frac{d\Phi_\neut}{dE_\neut\,d\cos\theta\,dX} \,e^{-\ell/\lambda_{\neut}}\,,
\label{eq:flux}
\end{equation}
where $\ell$ corresponds to the traveled distance of the neutralino, measured from the production point, which can be obtained from the height $h$ and the zenith angle $\theta$ using the geometrical relation
\begin{equation}
h= \sqrt{R_{\oplus}^{2} +2\ell R_{\oplus}\cos\theta + \ell^{2}} - R_{\oplus},    
\end{equation}
with $R_{\oplus}$ the earth's radius. 
In equation  (\ref{eq:flux}), the lifetime of the neutralino enters via the exponential decay factor that depends on the decay length $\lambda_{\neut}=\gamma_{\neut}\beta_{\neut}c\tau_{\neut}$.
\\

The neutralinos that reach the Super-Kamiokande detector can decay to SM particles inside its instrumental volume, which we model as a cylinder of 20 meters in radius and 40 meters in height \cite{Super-Kamiokande:2002weg}. Among the possible decay products of the neutralinos it is possible to find charged leptons and mesons, as well as neutral pseudoscalar and vector mesons. Different decay products correspond to different kinds of signals in the detector, which at Super-Kamiokande can be classified in two:  showering (or $e$-like), and non-showering (or $\mu$-like events). The former originates from electromagnetic and hadronic showers in the detector, while the latter is primarily associated to muons, which leave a distinctive Cherenkov ring with crisp edges. In this work, we focus on the showering signals originated from the decay of neutralinos described in table \ref{tab:benchmark_table}. In particular, for the case of neutral mesons in the final states, we assume that they decay to a showering signal inside the volume of the detector. The parameter choice of the benchmark B1 also allows $K^0_L$ as a final state for neutralino decays, but we do not consider this as part of our signal since this particle will typically decay outside the detector due to its large decay length.\\
For our benchmark scenarios, the important decay width formulas of neutralinos to pseudoscalar and vector mesons accompanied with a lepton are \cite{deVries:2015mfw}.
\begin{align}
\Gamma(\neut\rightarrow M_{jk} + \ell_i) &= \frac{\mathcal{K}^{1/2}(m^2_{\neut},m^2_{M_{jk}},m^2_{\ell_i})}{128\pi m^3_{\neut}(m_{q_j}+m_{q_k})^2}|G^{S}_{ijk}|^2f^2_{M_{jk}}\nonumber\\&\times m^4_{M_{ik}}(m^2_{M_{ab}} - m^2_{\neut} - m^2_{\ell_i}),\\[2ex]
\Gamma(\neut\rightarrow M^*_{jk} + \ell_i) &= \frac{\mathcal{K}^{1/2}(m^2_{\neut},m^2_{M^*_{jk}},m^2_{\ell_i})}{2\pi m^3_{\neut}}|G^{T}_{ijk}|^2\nonumber\\
&\times(f^V_{M^*_{jk}})^2\big[2(m^2_{\neut} - m^2_{\ell_i})^2 \nonumber\\
&\,\,\,- m^2_{M^*_{jk}}(m^2_{M^*_{jk}} + m^2_{\neut} + m^2_{\ell_i})\big],
\end{align}
where
\begin{align}
G^{T}_{ijk} \equiv \frac{g_2}{4\sqrt{2}}\frac{\lambda^{\prime}_{ijk}}{m^2_{\tilde{f}}}\tan\theta_W.\label{eq:tensor_vertex}
\end{align}
In the equation above, $M_{ij}$ ($M^*_{ij}$) represents one of the final state pseudoscalar (vector) mesons listed in the neutralino decays of table \ref{tab:benchmark_table}. The indices $i,j$ designate the family of the meson's valence quarks. For the final state pseudoscalar mesons, we use the decay constants $f_{\pi} \simeq 130$ MeV \cite{10.1093/ptep/ptaa104}, $f_{\pi^0} = f_{\pi}/\sqrt{2}$ and $f_{K^0} = f_{K}/\sqrt{2}$ \cite{Dreiner:2006gu}, where $f_{K}$ was defined after equation (\ref{eq:scalar_vertex}). For vector mesons, on the other hand, $f^V_{M^*}$ represents the vector meson decay constant, and for $K^{*0}$, $K^{*+}$ we use the approximate value $f^T_{K^{*}} \simeq 230$ MeV \cite{Dreiner:2006gu,deVries:2015mfw}.
\begin{table}[]
    \centering
    \begin{tabular}{||c||c||c||c||}
    \hline
        & \textbf{RPV coupling} & \textbf{Production} & \textbf{Decay mode} \\\hline \hline
        &  &  & \multicolumn{1}{l||}{$\neut\xrightarrow[]{\lambda'_{121}}K^0_{S} + \nu_e$}\\
        & & &\multicolumn{1}{l||}{$\neut\xrightarrow[]{\lambda'_{121}} K^{*0} +\nu_e$}\\
        \textbf{B1} & $\lambda'_{121},\lambda'_{112}$ & $D^\pm\xrightarrow[]{\lambda'_{121}}e^\pm + \neut$ &\multicolumn{1}{l||}{$\neut\xrightarrow[]{\lambda'_{112}} K^{(*)+} + e^-$}\\
        & & &\multicolumn{1}{l||}{$\neut\xrightarrow[]{\lambda'_{112}} K^0_{S} +\nu_e$}\\
        & & &\multicolumn{1}{l||}{$\neut\xrightarrow[]{\lambda'_{112}} K^{*0} +\nu_e$}\\
        \hline
        \multirow{2}{*}{\textbf{B2}} & \multirow{2}{*}{$\lambda'_{112},\lambda'_{111}$} & \multirow{2}{*}{$K^\pm\xrightarrow[]{\lambda'_{112}}e^\pm + \neut$} & \multicolumn{1}{l||}{$\neut\xrightarrow[]{\lambda'_{111}} \pi^+ + e^-$} \\
        & & &\multicolumn{1}{l||}{$\neut\xrightarrow[]{\lambda'_{111}} \pi^{0} + \nu_e$}\\\hline
    \end{tabular}
    \caption{Parameter choices that define our benchmark scenarios B1 and B2. The CP conjugate processes of all the decays shown are also allowed and therefore contribute to the neutralino decay width. The decay modes displayed in this table are compatible with a shower (or $e$-like) signal in Super-Kamiokande. See text for details.}
    \label{tab:benchmark_table}
\end{table}
As mentioned before, although for the production of neutralinos from $D$-mesons and kaons we only need  the trilinear couplings $\lambda^{\prime}_{121}$ and $\lambda^{\prime}_{112}$, the decay of neutralinos produced from kaons into visible showers in SK proceeds via the coupling $\lambda^{\prime}_{111}$. The two benchmark scenarios considered in this article can be seen in table \ref{tab:benchmark_table}. For benchmark B1 we consider neutralino masses in the range $m_{K^+} + m_{e}\leq m_\neut\leq m_{D^+} - m_e$, while for B2 the neutralino mass lies in the range $m_{\pi^0} + m_{e}\leq m_\neut\leq m_{K^+} - m_e$. The election of B1 allows for a direct comparison with references \cite{deVries:2015mfw,Dercks:2018eua}, where the sensitivity reach for future experiments aiming to explore the lifetime frontier was estimated. The election of B2 allows us to study a novel production channel where neutralinos can be abundantly produced in atmospheric showers. Furthermore, it can be seen from equations (\ref{eq:scalar_vertex}) and (\ref{eq:tensor_vertex}) that it is possible to combine the dependence of the decay widths on the RPV parameters and the sfermion mass in the ratio $\lambda^\prime_{ijk}/ m^2_{\tilde{f}}$, which we set as a free parameter.\\

\begin{figure}
\begin{center}
\includegraphics[width=1.09\columnwidth]{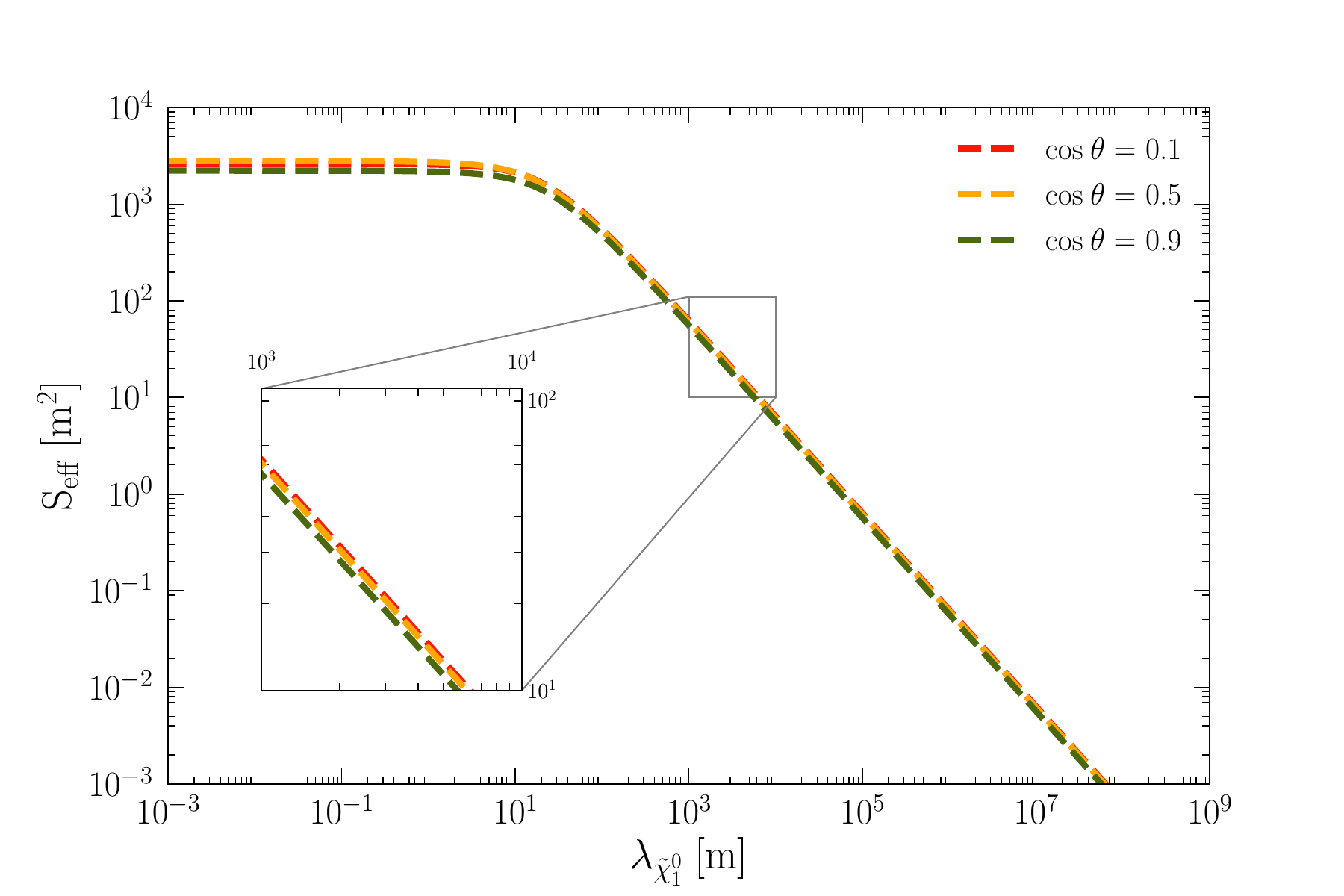}
\caption{
\textbf{\textit{Effective Surface for decay.}} The expected area of decay in Super-Kamiokande is shown as a function of the neutralino decay length for different incoming directions fixed by values of the cosine of the zenith angle at 0.1 (red line), 0.5 (yellow line), and 0.9 (green line). A similar result holds for negative values of $\cos\theta$. 
}
\label{fig:S_eff}
\end{center}
\end{figure}

To obtain the expected number of events with an energy in the range $E_\neut$ and $E_\neut + dE_\neut$, and with trajectories within $\cos\theta$ and $\cos\theta + d\cos\theta$, we include the effective surface $\mathcal{S}_{\text{eff}}$ for a decay to take place inside the SK detector, such that for a given time window $\Delta T$ the event rate is
\begin{equation}
\frac{dN}{dE_{\neut}\,d\cos\theta}= \Delta T \times \frac{d\Phi_\neut}{dE_\neut\,d\cos\theta} \times \mathcal{S}_{\text{eff}}, 
\label{eq:Rate}
\end{equation}
where the effective surface is obtained by integrating the surface of the detector that is perpendicular to the incoming direction of the neutralino flux, weighted by the probability that the particle decays inside the detector:
\begin{equation}
\mathcal{S}_{\text{eff}}= \int dS_{\perp}\,  \left(1 - e^{-\Delta\ell_{\text{det}}/\lambda_{\neut}}\right).
\end{equation}
Here, $\Delta \ell_{\text{det}}$ is the segment of the particle trajectory that traverses the detector, for which explicit analytical expressions can be found in the appendix of reference \cite{Arguelles:2019ziu}. The computation of the effective surface for decay is a purely geometrical problem. Figure \ref{fig:S_eff} shows the effective surface as a function of the decay length of the neutralino, with trajectories fixed by different values of the cosine of the zenith angle given by $0.9$, $0.5$ and $0.1$.
The event distribution in equation (\ref{eq:Rate}) can be integrated in energies and trajectories within a given bin determined by the resolution of the experiment. As mentioned before, we use atmospheric neutrino data reported by the Super-Kamiokande experiment in reference \cite{Abe_2018}. This data contains the angular distribution of events involving electron and muon neutrinos, from different energy regimes; the \textit{Sub-GeV} and \textit{Multi-GeV} sample of events with energies below and above 1330 MeV, respectively. Taking into account the trilinear couplings considered in this article, as well as the minimum energy available in \texttt{MCEq}, we chose the showering (or $e$-like) event sample in the \textit{Multi-GeV} energy window. The total events reported in reference \cite{Abe_2018} correspond to the SK-I up to the SK-IV data taking periods with a total run of 5,326 days.\\
\begin{figure}[hbt!]
\centering
\includegraphics[width=0.99\columnwidth]{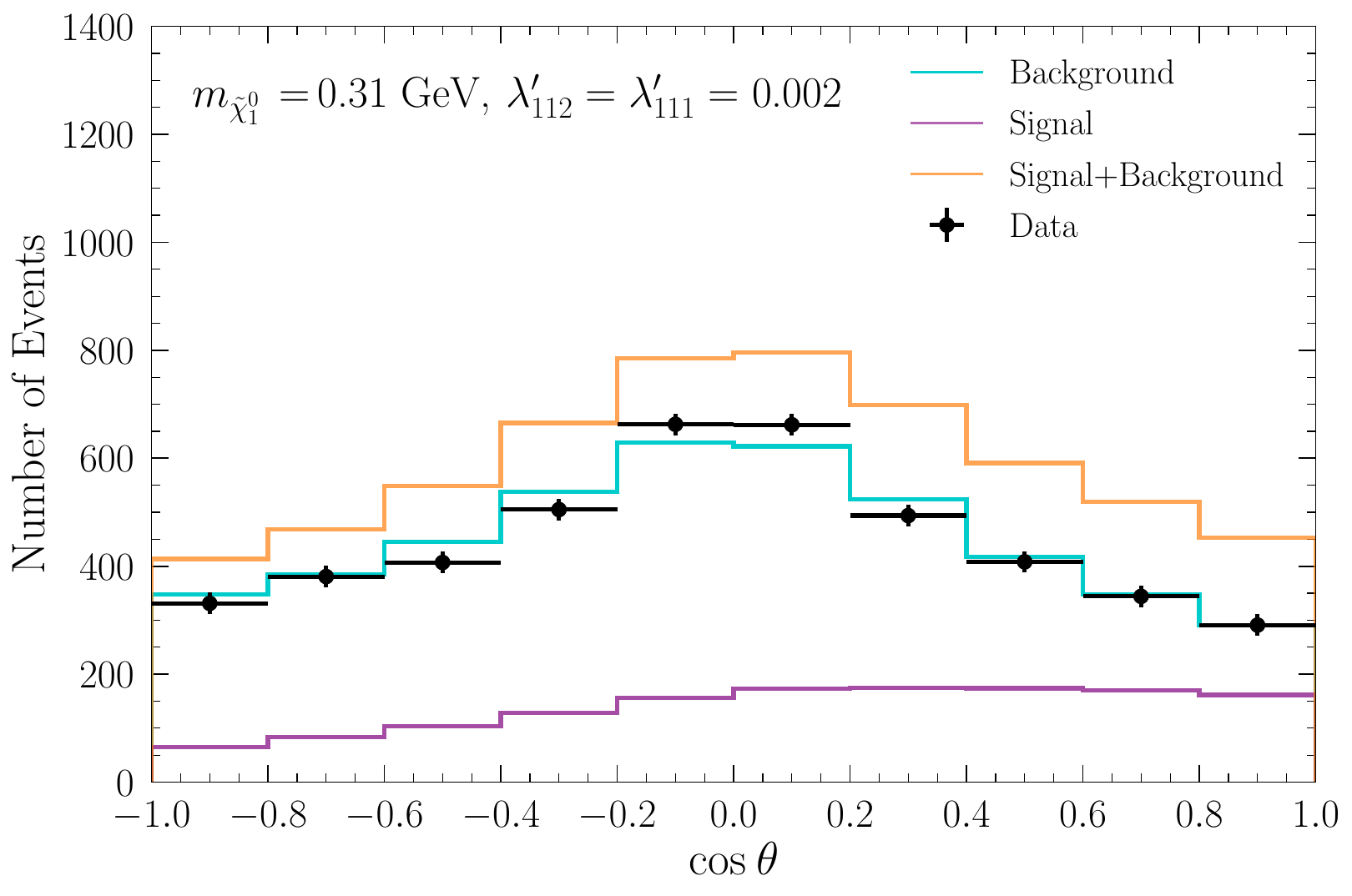} 
\caption{\textbf{\textit{Expected events at SK.} }
Distribution of atmospheric events in Super-Kamiokande for incoming directions fixed by the cosine of the zenith angle. Data points are shown in black, while the background is depicted by the cyan line. The signal produced from the decay of neutralinos generated from kaons is shown in purple. The signal plus background contribution is shown in orange. The mass value and RPV couplings are fixed as indicated in the figure. }
\label{fig:Events}
\end{figure}
\begin{figure*}[hbt!]
\begin{center}
\includegraphics[width=\linewidth]{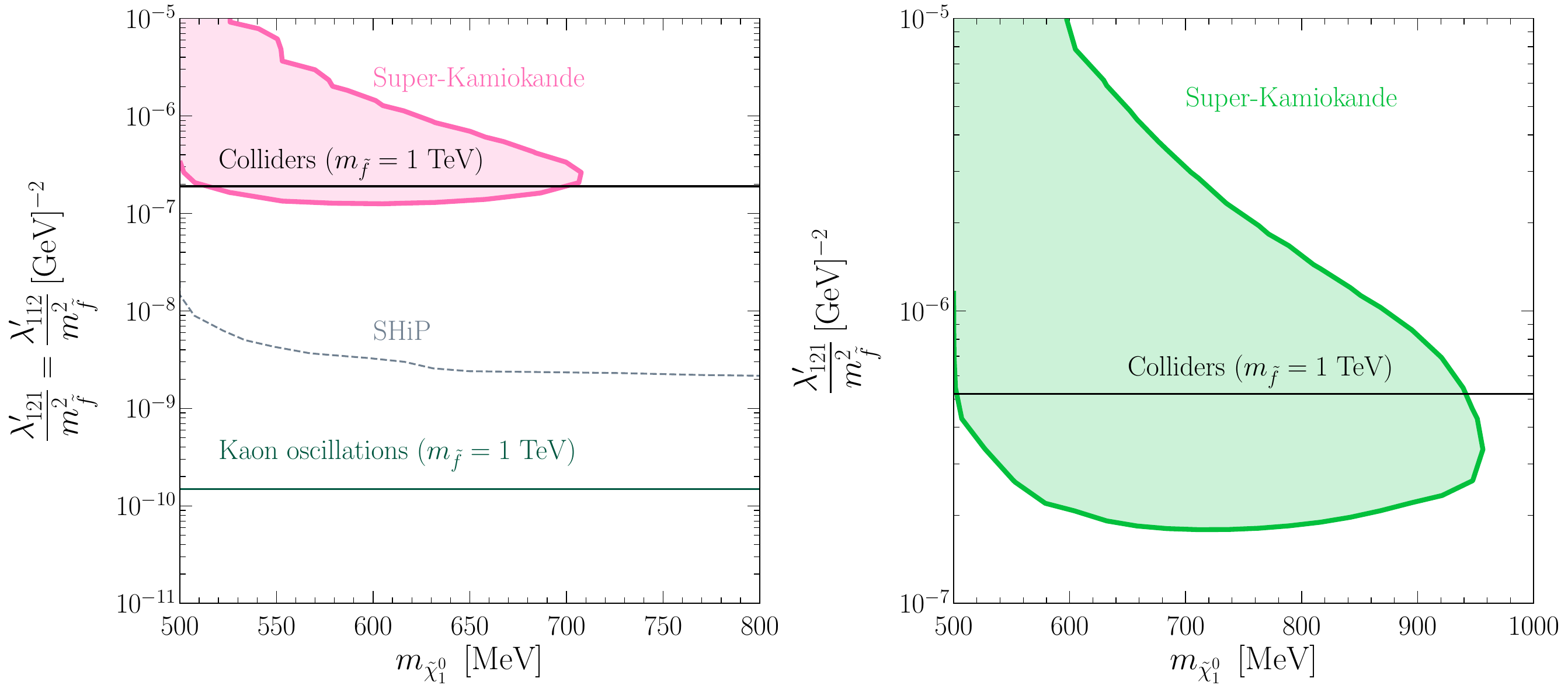}
\caption{
\textbf{\textit{Limits for benchmark B1.}}
Limits derived at 90\% confidence level in the parameter space defined by the neutralino mass $m_\neut$ and the trilinear RPV parameters $\lambda'_{121}$, $\lambda'_{112}$. On the left panel, the pink contour indicates the parameter space excluded by SK when the two RPV couplings are set to the same value, while the dashed line shows the projected SHiP sensitivity reach \cite{deVries:2015mfw}. On the right panel, the green contour indicates the excluded parameter space when $\lambda'_{121} \neq 0$ and $\lambda'_{112} = 0$. Solid lines indicate constraints from neutral kaon oscillations \cite{Domingo:2018qfg} in green and DY processes at the LHC (labeled as `Colliders') \cite{Bansal:2018dge} in black, both evaluated at a sfermion mass of $1$ TeV (for details, see text).}
\label{fig:D-meson-limit}
\end{center}
\end{figure*}
The number of events contained in the \textit{Multi-GeV} range for the \textit{i}-th bin in the cosine of the zenith angle is computed with the formula
\begin{align}
N_{c_i}=\, & \textrm{Br}(\neut \rightarrow e\textrm{-like}) \int_{\cos\theta_{i}-0.1}^{\cos\theta_{i}+0.1} d\cos\theta \nonumber\\ 
& \times\int_{E_{\text{min}}}^{E_{\text{max}}} dE_{\neut}\, \epsilon \frac{dN}{dE_{\neut}\,d\cos\theta},
\label{eq:Signal}
\end{align}
where $\epsilon$ is the detection efficiency, which for the \textit{Multi-GeV} sample is flat in energy with a value of $0.75$, while $ E_{\rm min}$ and $E_{\rm max}$ are equal to 1.5 and 90.5 GeV, respectively \cite{Abe_2018}. The background for our search corresponds to the expected $e$-like events from electron neutrinos at SK as reported in reference \cite{Abe_2018}. As an illustration, in figure \ref{fig:Events} we show an example of the expected event signal generated from neutralino decays, when the couplings $\lambda_{112}^{\prime}$ and $\lambda_{111}^{\prime}$ are both fixed to a value of $0.002$, and the neutralino mass is $0.31$ GeV.


\section{Results and Discussion} \label{sec:results}

\begin{figure*}[hbt!]
\begin{center}
\includegraphics[width=1\linewidth]{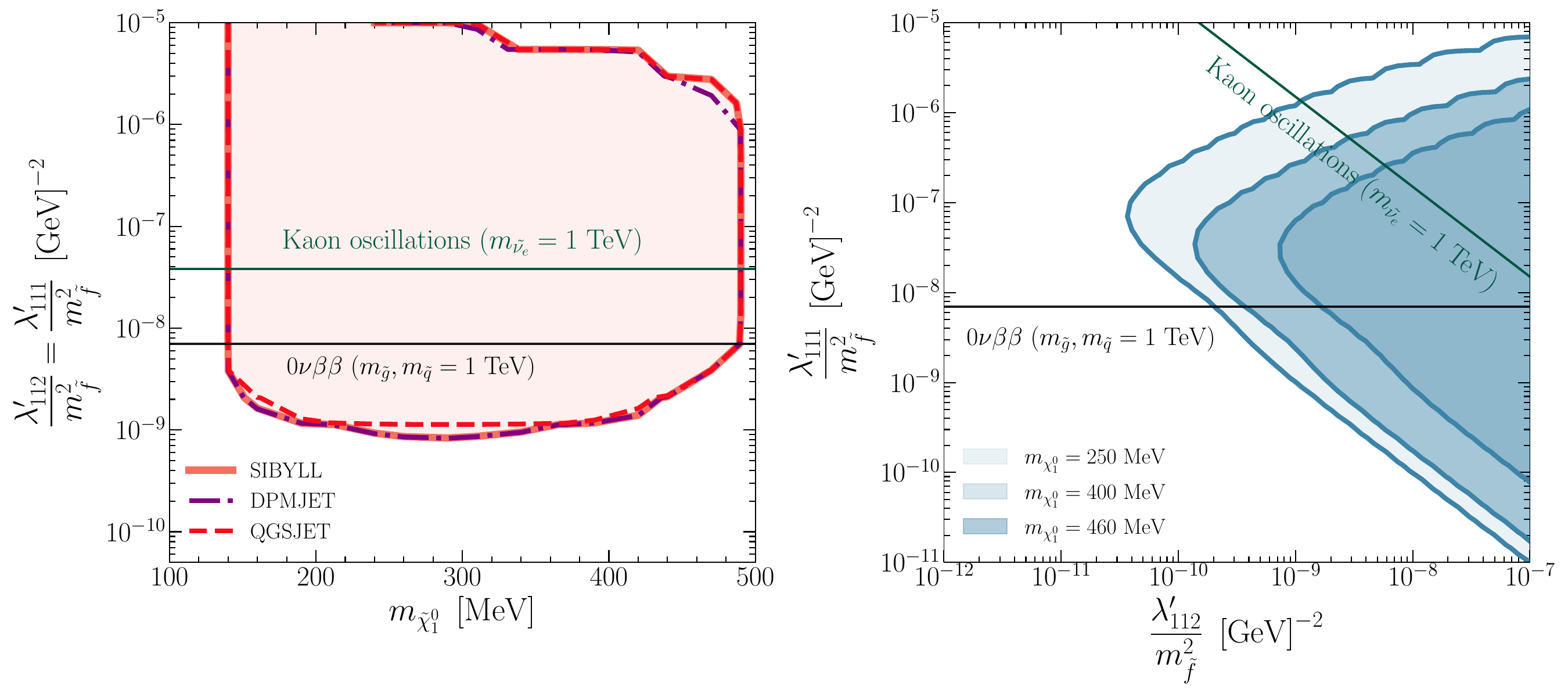}
\caption{
\textbf{\textit{Limits for benchmark B2.}} Limits derived at 90\% confidence level in the parameter space defined by the neutralino mass $m_\neut$ and the trilinear RPV parameters $\lambda'_{112}$, $\lambda'_{111}$. On the left panel, the red contours indicate the parameter space excluded by SK when the two RPV parameters included in B2 have the same value. The different types of lines are used to distinguish the contours obtained with three different event generators, \texttt{SYBILL-2.3}, \texttt{QGSJET-II-04}, and \texttt{DPMJET-III}. On the right panel, the \texttt{SYBILL-2.3} hadronic interaction model was used to generate three blue contours that indicate the parameter space excluded by SK when both RPV parameters vary freely. The different shades of blue represent different values of the neutralino mass. Solid lines indicate constraints from kaon oscillations \cite{Domingo:2018qfg} in green and neutrinoless double beta decay ($0\nu\beta\beta$) \cite{Deppisch:2020ztt} in black. The bounds are evaluated for degenerated sfermion masses set to a value of 1 TeV.}
\label{fig:Kaon-limit}
\end{center}
\end{figure*}

The signal computed in equation (\ref{eq:Signal}) depends on the neutralino mass and RPV couplings through its lifetime, branching fraction of production from mesons, and the fraction of neutralinos that decay to an $e$-like signal in the detector.
We adopt the distribution for Poisson events, which implies a chi-squared statistic of the form
\begin{align}
\chi^2 =\,&2  \sum_{c_i}\bigg[  N_{c_i}(m_{\neut},\lambda^{\prime})+ B_{c_i} - D_{c_i}\nonumber\\
&+D_{c_i} \log\left({\frac{D_{c_i}}{N_{c_i}(m_{\neut},\lambda^{\prime})+ B_{c_i}}}\right)\bigg],
\label{eq:chi2}
\end{align}
where the sum runs over the 10 bins in cosine of the zenith angle, and $B_{c_i}$, $D_{c_i}$ are the background and data in the \textit{i}-th bin, respectively. We apply this statistical test to the SK data and derive constraints at 90$\%$ confidence level. The results for our two benchmark scenarios are displayed in figures \ref{fig:D-meson-limit} and \ref{fig:Kaon-limit}.\\
There are numerous phenomenological constraints on the lepton number violating operator $\lambda'_{ijk} \wh{L}_i \wh{Q}_j \wh{D}^c_k$ (see \cite{Dreiner:2006gu} and references therein). In particular, as pointed out in \cite{deCarlos:1996yh}, RPV parameters can be subject to strong constraints due to their contribution to meson oscillation observables. This is the strongest bound that applies to the benchmark B1 specified in table \ref{tab:benchmark_table}, where the tree-level contributions to kaon oscillations induced by that choice of parameters imply the sneutrino mass dependent bound \cite{Domingo:2018qfg}
\begin{align}
|\lambda'_{112}\lambda'_{121}|\leq2.2\times10^{-8}\left(\frac{m_{\tilde{\nu}_e}}{1\text{ TeV}}\right)^2,\label{eq:bounds_112_vs_121}
\end{align}
where $m_{\tilde{\nu}_e}$ is the sneutrino mass. Our results for B1 are shown figure \ref{fig:D-meson-limit}, where we evaluated the limits for a sneutrino and squark mass of 1 TeV. As it can be seen on the left panel, the limits imposed by kaon oscillations for degenerated sneutrino and squark masses exclude all the parameter space within the projected sensitivity reach at SHiP \cite{deVries:2015mfw}, our limits from SK, and the limit from the Drell-Yan (DY) monolepton process $pp\rightarrow\ell\overline{\nu}$  at the LHC
(labeled as `Colliders'). The latter is a single coupling bound for the parameter $\lambda'_{112}$ given by \cite{Bansal:2018dge}
\begin{align}
\lambda'_{112}\leq 0.16\frac{m_{\tilde{s}_R}}{1\text{ TeV}} + 0.030,\label{eq:bounds_112}
\end{align}
where $m_{\tilde{s}_R}$ is the strange squark mass, which is also set to 1 TeV to compare with our results. However, since in Super-Kamiokande all the neutralino decay products listed in table \ref{tab:benchmark_table} are visible as $e$-like events, we can still have a signal when the parameter $\lambda'_{112}$ is set to zero and we are left with neutral kaons in the final states. In such a case, the limits from kaon oscillations do not apply (see figure \ref{fig:D-meson-limit}, right panel), and we are left with the collider constraint from the DY dilepton process $pp\rightarrow \ell^+\ell^-$ for the parameter $\lambda'_{121}$, which is given by \cite{Bansal:2018dge}.
\begin{align}
\lambda'_{121}\leq 0.34\frac{m_{\tilde{q}}}{1\text{ TeV}} + 0.18,\label{eq:bounds_121}
\end{align}
where $m_{\tilde{q}}$ is the squark mass. In this case, we find that the limit obtained from Super-Kamiokande data is better than the current constraint for the corresponding mass range.\\
\begin{figure*}[hbt!]
\centering
\includegraphics[width=\linewidth]{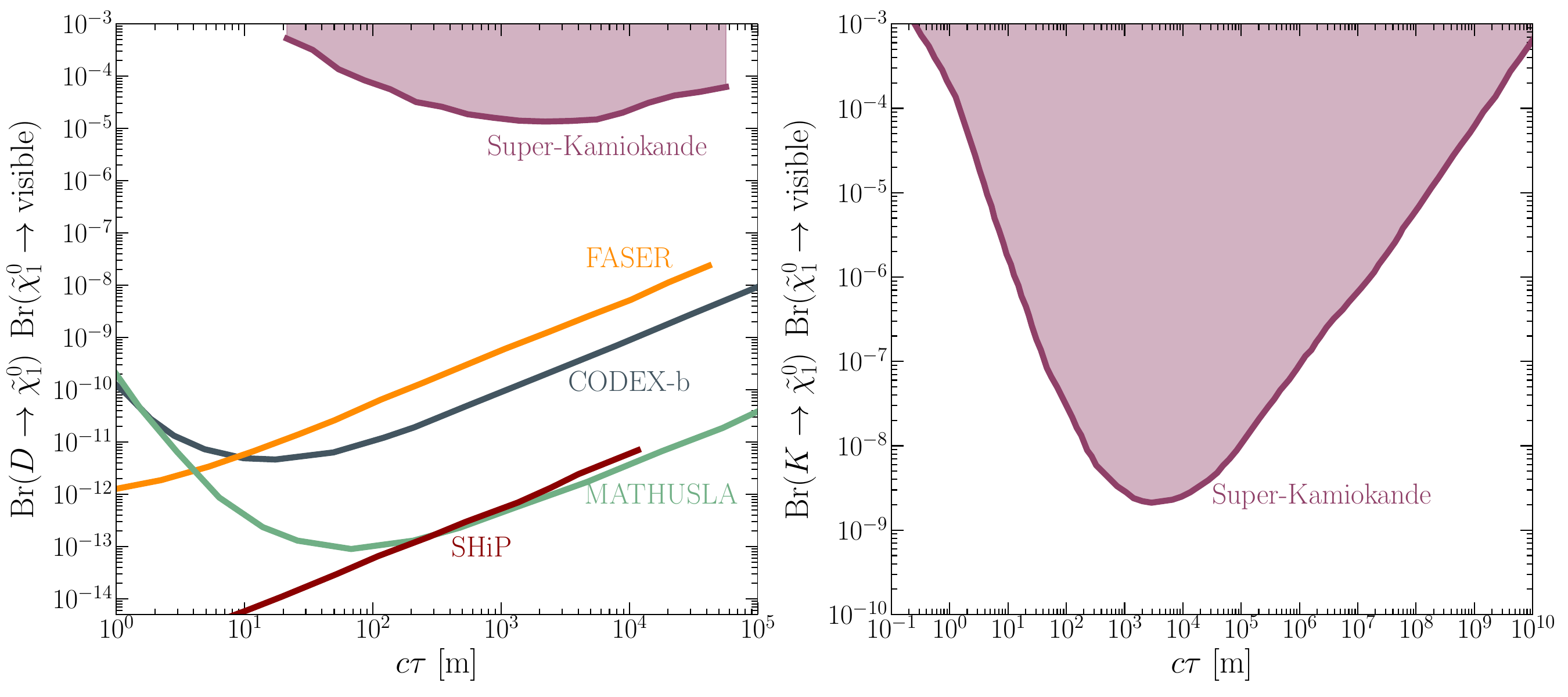} 
\caption{\textbf{\textit{Comparison with lifetime frontier experiments.}} Limits at 90\% confidence level in the plane defined by the total branching ratio and the proper decay length of the long-lived neutralino. On the left panel, the purple contour shows the limits for SK derived in this article for benchmark B1, and the coloured lines displays the projected sensitivity reach at future experiments for the same benchmark scenario derived in references \cite{deVries:2015mfw,Dercks:2018eua}. On the right panel, the limit obtained using Super-Kamiokande data for the benchmark scenario B2 is shown. There are no other studies for this case in the literature.}
\label{fig:LifetimeLimit}
\end{figure*}
On the other hand, in the case of benchmark B2, there is also a stringent limit from kaon oscillations on the product of the parameters, namely \cite{Domingo:2018qfg} 
\begin{align}
|\lambda'_{111}\lambda'_{112}|\leq1.5\times10^{-3}\left(\frac{m_{\tilde{\nu}_e}}{1\text{ TeV}}\right)^2.\label{eq:bounds_111_vs_112}
\end{align}
Nevertheless, the most stringent constraint comes from its contribution to $0\nu\beta\beta$.
The current limit for the half life of this process as determined by the KamLAND-Zen experiment \cite{KamLAND-Zen:2016pfg}, imposes the upper limit \cite{Deppisch:2020ztt}
\begin{align}
\lambda'_{111}\leq2.2\times10^{-3}\left(\frac{m_{\tilde{q}}}{1\text{ TeV}}\right)^2\left(\frac{m_{\tilde{g}}}{1\text{ TeV}}\right)^{1/2},\label{eq:0nuBB_limit}
\end{align}
where $m_{\tilde{q}}$ and $m_{\tilde{g}}$ are the masses of the squarks and gluinos, respectively. To contrast these bounds with our results, again we set both mass parameters to the value of 1 TeV (see figure \ref{fig:Kaon-limit}). Remarkably enough, due to the higher kaon flux with respect to $D$-mesons, the limits from Super-Kamiokande in this parameter space turn out to be more stringent than the bounds that come from both kaon oscillations and $0\nu\beta\beta$. The excluded parameter space for this benchmark scenario changes marginally with different choices of hadronic interaction models, as it can be seen from the lines that indicate the exclusion region obtained with each model (see the left panel of figure \ref{fig:Kaon-limit}). The difference between the kaon flux obtained with \texttt{EPOS-LHC} and with our benchmark model \texttt{SYBILL-2.3} does not translate to any visible impact in the excluded parameter space, therefore we omit its contour in the figure. As a caveat, we note that the bound shown in equation (\ref{eq:0nuBB_limit}) relies upon the \textit{gluino dominance} assumption, where the neutralino mass is a fraction of the order of $10^{-2}$ times the gluino mass \cite{Hirsch:1995ek}.\\
Finally, we emphasize that our results do not depend on the value of the sfermion masses. However, the bounds shown in equations (\ref{eq:bounds_112_vs_121}--\ref{eq:0nuBB_limit}) become more strict as sfermion masses increase and less strict when they are lowered. Moreover, in the case where the squarks and sneutrino masses are kept fixed and the gluino mass is increased, the bound from $0\nu\beta\beta$ becomes less stringent.\\

The excluded combinations of values for the neutralino masses and RPV couplings allow us to determine the region of the parameter space that can be probed in the plane defined by the total branching fraction, which is the production branching ratio of the neutralino times the branching ratio of its decay to a visible signal, and the lifetime of the neutralino. The results can be seen in figure \ref{fig:LifetimeLimit}. Given that for benchmark B1 we are considering the same production and decay channels as in references \cite{deVries:2015mfw} and \cite{Dercks:2018eua}, we can compare the current region that can be excluded by Super-Kamiokande, with the sensitivity reach expected for FASER, and other possible future experiments including  CODEX-b, MATHUSLA and SHiP (left panel of figure \ref{fig:LifetimeLimit}).\\

In the case where the neutralinos are produced in kaon decays, there are no studies on the expected sensitivity reach at the next generation of detectors. The results obtained in this case demonstrate the great capacity of the Super-Kamiokande neutrino detector to place stringent limits for long-lived particles, owed to its ability to probe particles with lifetimes with a peak sensitivity around 10 km and total branching fractions of order $10^{-9}$.

\section{Conclusions} \label{sec:conclusion}  

The lifetime frontier has emerged as a powerful line of exploration to search for beyond the Standard Model physics, specially in the absence of new signals at the LHC. R-parity-violating supersymmetry with light, long-lived neutralinos constitutes a well-motivated scenario for new physics, with a rich phenomenology that has been gathering attention in recent years. In this work, we demonstrate a new way to search for long-lived neutralinos with masses of order $0.1-1$ GeV that could be produced from the decay of charged $D$-mesons and kaons in cosmic-ray air showers, and whose visible decay can take place within large neutrino detectors such as Super-Kamiokande. We have analyzed two benchmark scenarios that include the couplings $\lambda_{121}^{\prime}$,  $\lambda_{112}^{\prime}$,  and $\lambda_{111}^{\prime}$. In both cases, it is possible to improve the excluded region in parameter space when compared to existing constraints from colliders and neutrinoless double beta decay. Note, however, that this comparison requires to fix the mass of the sfermions, which we set to 1 TeV. An interesting feature about benchmark B1 is that in this case it is possible to compare the lifetime range that can be probed using SK data, against the expected sensitivity reach of next-generation, long-lived particle detectors. We find that the sensitivity for Super-Kamiokande peaks for lifetimes of the order of 1.0 km. In the case of the benchmark scenario B2, the advantage is twofold. On the one hand, there are currently no searches for light neutralinos produced from kaon decays. On the other hand, since kaons have a considerably higher production rate in air showers (as it can be inferred from figure \ref{fig:neutralino_production}), we find that it is possible to probe neutralinos with lifetimes of the order of hundreds of kilometers, while also being able to achieve a limit in $\lambda^\prime_{111}/m^2_{\tilde{f}}$ and $\lambda^\prime_{112}/m^2_{\tilde{f}}$ of order $10^{-9}$ GeV$^{-2}$,  assuming these RPV couplings are equal.\\
When possible, we quantified the uncertainty that results from the choice of different hadronic interaction models. This was not possible for the benchmark scenario B1, since \texttt{SYBILL-2.3} is the only event generator that provides a state-of-the-art simulation of charmed mesons. For the case of the benchmark scenario B2, we find no strong dependence on the hadronic interaction model chosen to simulate the production of kaons in the atmospheric shower.

Overall, the results found in this study demonstrate the potential of large neutrino detectors to place limits on beyond the SM scenarios predicting long-lived particles, specially when contrasted with supersymmetric searches at colliders, where signals must be carefully chosen, and the reinterpretation of results is usually a complicated task. Finally, we stress that further scenarios can be pursued systematically along the direction presented in this work.\\
\acknowledgments
We acknowledge Jordy de Vries, Zeren Simon Wang and Anatoli Fedynitch for helpful discussions. The work of V.M. is supported by ANID-PCHA/DOCTORADO BECAS CHILE/2018-72180000. The work of P.C. is funded by Becas Chile, ANID-PCHA/2018-72190359. G.C. acknowledges support from ANID FONDECYT-Chile grant No. 3190051. The Feynman diagrams shown in
this article were generated with the TikZ-Feynman package \cite{Ellis:2016jkw}.
\bibliography{bibliography}
\end{document}